\documentclass[12pt]{article}

\usepackage{amsmath,amssymb}
\usepackage{amsfonts}
\usepackage{graphicx}
\usepackage{a4wide}

\usepackage[T2A]{fontenc}
\usepackage[cp1251]{inputenc}
\usepackage[russian]{babel}

\begin{document}

\noindent {\it Problems of Information Transmission},\\
\noindent vol. 50, no. 3, pp. 19--34, 2014.

\begin{center} {\bf M. V. Burnashev\footnote[1]
{Supported in part by the Russian Foundation
for Basic Research, project nos.~12-01-00905a and 13-01-12458 ofi\_m2.},
H. Yamamoto\footnote[2]
{Supported in part by the Japanese Fund of JSPS KAKENHI, grant
no.~25289111.}} \end{center}

\vskip 0.4cm

\begin{center}
{\large\bf ON USING FEEDBACK IN A GAUSSIAN CHANNEL}
\end{center}

{\begin{quotation} \normalsize For information transmission a
discrete time channel with independent \\
additive Gaussian noise is used. There is also another channel with
independent additive Gaussian noise (the feedback channel), and the
transmitter observes without delay all outputs of the forward channel
via that channel. Transmission of nonexponential number of messages
is considered (i.e. transmission rate equals zero) and the achievable
decoding error exponent for such a combination of channels is
investigated. The transmission method strengthens the method used by
authors earlier for BSC and Gaussian channels. In particular, for
small \\
feedback noise, it allows to gain 33.3\% (instead of 23.6\%
earlier in the similar case of Gaussian channel).

\end{quotation}}

\vskip 0.7cm

\begin{center}
{\large\bf \S\,1. Introduction and main result}
\end{center}

In the paper results of \cite{BY2010} are strengthened and proofs
are simplified. We consider the discrete time channel with
independent additive Gaussian noise, i.e. if
$\mbox{\boldmath $x$} = (x_1,\ldots,x_n)$ is the input codeword
then the received block $\mbox{\boldmath $y$} = (y_1,\ldots,y_n)$ is
\begin{equation}\label{defchan1}
y_{i} = x_{i} + \xi_{i}, \qquad i=1,\ldots,n,
\end{equation}
where $\mbox{\boldmath $\xi$} = (\xi_{1},\ldots,\xi_{n})$ are
independent ${\cal N}(0,1)$--Gaussian random variables, i.e.
${\mathbf E} \xi_{i} = 0,\; {\mathbf E} \xi_{i}^2 = 1$. There is the
noisy feedback channel, and the transmitter observes (without delay)
all outputs $\{z_{i}\}$ of the forward channel via that noisy
feedback channel
\begin{equation}\label{deffeed1}
z_{i} = y_{i} + \sigma\eta_{i}, \qquad i=1,\ldots,n,
\end{equation}
where $\mbox{\boldmath $\eta$} = (\eta_{1},\ldots,\eta_{n})$ are
independent (and independent of $\xi$) ${\cal N}(0,1)$--Gaussian
random variables, i.e.
${\mathbf E} \eta_{i} = 0,\; {\mathbf E}\eta_{i}^2 = 1$. The value
$\sigma > 0$, characterizing feedback channel noise intensity, is
given. No coding is used in the feedback channel (i.e. the receiver
simply re-transmits all received outputs to the transmitter). In
other words, the feedback channel is ``passive''.

We assume that the input block $\mbox{\boldmath $x$}$ satisfies the
constraint
\begin{equation}\label{constr1}
\sum_{i=1}^{n}x_{i}^{2} \leq nA,
\end{equation}
where $A$ is a given constant. We denote by AWGN$(A)$ the channel
(\ref{defchan1}) with constraint (\ref{constr1}) without feedback,
and by AWGN$(A,\sigma)$ that channel with noisy feedback
(\ref{deffeed1}). The capacity of both channels equals
$C(A) = [\ln(1+A)]/2$.

We consider the case when the overall transmission time $n$ and
$M = e^{o(n)}$, $n \to \infty$, equiprobable messages
$\{\theta_{1},\ldots,\theta_{M}\}$ are given. After the moment
$n$, the receiver makes a decision ${\hat \theta}$ on the message
transmitted. We are interested in the best possible decoding error
exponent (and whether it exceeds the similar exponent of the channel
without feedback).

It is well known \cite{Shannon56} that even
noiseless feedback does not increase the capacity of the Gaussian
channel (or any other memoryless channel). However, feedback allows
to improve the decoding error exponent
(\emph{channel reliability function})
with respect to no-feedback \\ channel. Possibility of such
improvement stimulated a good interest to that topic in 60--80's.
A good number of interesting results have been obtained during that
period (e.g. [3--10]). Unfortunately, all those papers had a common
drawback:  their methods were heavily based on the assumption that
the feedback is {\it noiseless}. It was necessary in order to have
perfect mutual coordination between both the transmitter and the
receiver. Essentially, any noise in the feedback link destroyed that
coordination and all hypothetical improvements. It was not clear
whether it is possible to improve communication characteristics
using more realistic noisy feedback.

That uncertainty with noisy feedback remained till 2008, when in
\cite{BY0}--\cite{BY2} it was shown (for BSC) how to use such
feedback in order to improve the decoding error exponent. Although
improvement was not large (approximately 14.3\%. for small feedback
noise), it was the first method that worked for noisy feedback. Later
results (\cite{BY2010} and this paper) are developments of
\cite{BY0}--\cite{BY2}.

In order to explain what is new in the paper, remind briefly what
was done in earlier papers  \cite{BY0}--\cite{BY2} and \cite{BY2010}.
For that purpose we explain first why noiseless feedback allows to
improve decoding error exponent. For a channel without feedback that
exponent is determined (for small transmission rates $R$) by the code
distance of the code used (i.e. by the minimal distance among
codewords). Noiseless feedback allows during transmission to change
the code (code function) used, e.g. increasing the distances among
most probable codewords. That feature allowed to improve the
decoding error exponent. But for that purpose an ideal coordination
between both the transmitter and the receiver are required.

In all papers \cite{BY0}--\cite{BY2}, \cite{BY2010} and this one
coding function can be changed only at one fixed moment
(``switching moment''). In \cite{BY0}--\cite{BY2} such change took
place only if two most probable codewords were much more probable
than all remaining codewords. It was shown that if noise in the
feedback channel is less than a certain critical value
$p_{\rm crit}$, then it is possible to choose transmission
parameters such that the probability of miscoordination between
the transmitter and the receiver becomes smaller than decoding
error probability. That fact allowed to improve the decoding error
exponent with respect to no-feedback channel.

Later in the paper \cite{BY2010} for Gaussian channel that method
was strengthened taking into account not two, but three most probable
codewords. Moreover, the decoding method was improved. It allowed
not only to improve the gain (23.6\% instead of 14.3\% in \cite{BY1},
but also to show that for any {\it noise intensity}
$\sigma^{2} < \infty$ it is possible to improve the best error
exponent of AWGN$(A)$ no-feedback channel. Of course, if $\sigma$ is
not small then the gain is small, but it is strongly positive.
In other words, in the problem considered there is no any critical
level $\sigma_{\rm crit}$, beyond which it is not possible to
improve the error exponent of the no-feedback channel. It should be
noticed also that the investigation method with optimal decoding in
\cite{BY2010} was rather tedious.

The method of papers \cite{BY0,BY1} was applied to Gaussian channel
AWGN$(A,\sigma)$ in \cite{XiKim1} with similar to \cite{BY0,BY1}
results (in particular, with the same asymptotic gain 14.3\%).

The aim of the paper is to strengthen the transmission method
\cite{BY2010} (in particular, using up to four most probable
codewords) and also simplify its analysis. It allows
to improve the gain up to 33.3\% (instead of 23.6\% in
\cite{BY2010}).

{\it Remark} 1. We consider the case when the value $\sigma^{2} > 0$
is fixed and does not depend on the number of messages $M$.

For $\mbox{\boldmath $x$},\mbox{\boldmath $y$} \in
{\mathbb{R}}^{n}$ denote
$$
(\mbox{\boldmath $x$},\mbox{\boldmath $y$}) =
\sum\limits_{i=1}^n x_{i} y_{i}, \quad \|\mbox{\boldmath $x$}\|^2 =
(\mbox{\boldmath $x$},\mbox{\boldmath $x$}), \quad
d\left(\mbox{\boldmath $x$},\mbox{\boldmath $y$}\right) =
\|\mbox{\boldmath $x$} - \mbox{\boldmath $y$}\|^{2}.
$$
A subset ${\cal C} = \{\mbox{\boldmath $x$}_{1},
\ldots, \mbox{\boldmath $x$}_{M}\}$ with
$\|\mbox{\boldmath $x$}_{i}\|^{2} = An$, $i = 1,\ldots,M$ is
called a $(M,A,n)$--code of length $n$.

For a code ${\cal C} = \{\mbox{\boldmath $x$}_{i}\}$ denote by
$P_{\rm e}({\cal C})$ the minimal possible decoding error
probability
$$
P_{\rm e}({\cal C}) = \min \max_{i}P(e|\mbox{\boldmath $x$}_{i}),
$$
where $P(e|\mbox{\boldmath $x$}_{i})$ -- conditional decoding error
probability provided $\mbox{\boldmath $x$}_{i}$ was transmitted,
and minimum is taken over all decoding methods (it will be
convenient to denote the message transmitted as $\theta_{i}$ and
$\mbox{\boldmath $x$}_{i}$ as well).

In the paper we consider the case when $M = M_{n} \to \infty$,
but $M_{n} = e^{o(n)}$ as $n \to \infty$ (it corresponds to zero-rate
of transmission). For $M$ messages and AWGN$(A)$ channel
denote by $P_{\rm e}(M,A,n)$ the minimal possible decoding error
probability for the best $(M,A,n)$--code and introduce the exponent
(in $n$) of that function \cite{Sh}
\begin{equation}\label{main2}
\begin{gathered}
E(A) = \limsup_{\substack{n \to \infty  \\ M \to \infty \\
\ln M = o(n)}}\,\frac{1}{n}\,\ln \frac{1}{P_{\rm e}(M,A,n)} =
\frac{A}{4}.
\end{gathered}
\end{equation}

Similarly, for AWGN$(A,\sigma)$ channel with noisy feedback denote by
$P_{\rm e}(M,A,\sigma,n)$ the minimal possible decoding error
probability and introduce the function
$$
F(A,\sigma) = \limsup_{\substack{n \to \infty  \\ M \to \infty \\
\ln M = o(n)}}\,\frac{1}{n}\,\ln\frac{1}{P_{\rm e}(M,A,\sigma,n)}.
$$

It is also known that if $\sigma = 0$ (i.e. noiseless feedback) then
\cite{Pin1}
\begin{equation}\label{main21}
\begin{gathered}
F(A,0) =  \frac{A}{2}.
\end{gathered}
\end{equation}

For AWGN$(A,\sigma)$ channel denote by $F_{1}(A,\sigma)$ the best
error exponent for the transmission method with one switching moment,
described in \S 2. Then
$F_{1}(A,\sigma) \leq F(A,\sigma)$ for all $A,\sigma$.

The paper main result is as follows.

{T h e o r e m}. {\it Let $M \to \infty$ and
$\ln M = o(n)$, $n \to \infty$. Then
the formula holds}
\begin{equation}\label{Theor1a}
\begin{gathered}
F_{1}(A,\sigma) \geq \frac{A(1 - \sigma^{2})}{3}.
\end{gathered}
\end{equation}

For small $\sigma$ the formula (\ref{Theor1a}) gives 33.3\% of
improvement with respect to no-feedback channel (see \eqref{main2}).
It is given in a simplified form oriented to small values of
$\sigma$. A more general formula (following from results of \S 4)
would be too bulky.

{\it Remark} 2.
The method described in the paper and its analysis can be
generalized on slow growing number $N = N(\sigma)$ of switches. It
allows to prove the following result
\begin{equation}\label{Theor1b}
\begin{gathered}
F_{N(\sigma)}(A,\sigma) = \frac{A(1 + o(\sigma))}{2}, \qquad
\sigma \to 0.
\end{gathered}
\end{equation}
In other words, for small $\sigma$ the formula (\ref{Theor1b})
gives improvement of 100\% with respect to no-feedback channel
(see \eqref{main2}), and it coincides with similar result
\eqref {main21} for noiseless feedback. It will be done in another
paper.

In \S 2 the transmission method with one switching moment and its
decoding are described. In \S\S\,3-4 its analysis is performed and
the theorem is proved. Greek letters $\xi,\eta,\zeta,\xi_{1},\ldots$
in the paper designate ${\cal N}(0,1)$--Gaussian random variables.

\begin{center}
{\large\bf \S\,2. Transmission/decoding method}
\end{center}

We use the transmission strategy with one fixed switching moment at
which the code used will be changed. Denote $n_{1} = n/2$ and
partition the total transmission time $[1,n]$ on two phases:
$[1,n_{1}]$ (phase I) and $[n_{1}+1,n]$ (phase II).
After moment $n$ the receiver makes a decision in favor
of the most probable message $\theta_{i}$ (based on all received on
$[1,n]$ signals).

Each of $M$ codewords $\{\mbox{\boldmath $x$}_{i}\}$ of length
$n$ have the form $\mbox{\boldmath $x$}_{i} =
(\mbox{\boldmath $x$}_{i}',\mbox{\boldmath $x$}_{i}'')$, where both
$\mbox{\boldmath $x$}_{i}'$ (to be used on phase I) and
$\mbox{\boldmath $x$}_{i}''$ (to be used on phase II) have length
$n_{1}$ .

Similarly, the received block $\mbox{\boldmath $y$}$
has the form $\mbox{\boldmath $y$} =
(\mbox{\boldmath $y$}',\mbox{\boldmath $y$}'')$, where
$\mbox{\boldmath $y$}'$ is the block received on phase I and
$\mbox{\boldmath $y$}''$ is the block received on phase II. Denote
by $\mbox{\boldmath $z$}'$ the received (by the transmitter) block
on phase I. The codewords first parts
$\{\mbox{\boldmath $x$}_{i}'\}$ are fixed, while the second parts
$\{\mbox{\boldmath $x$}_{i}''\}$ will depend on the block
$\mbox{\boldmath $z$}'$ received by the transmitter on phase I.

We set two positive constants $A_{1}, A_{2}$ such that
\begin{equation}\label{A1A2II}
A_{1} + A_{2} = nA,
\end{equation}
and denote
\begin{equation}\label{defbeta}
\beta = \frac{A_{2}}{A_{1}}.
\end{equation}
Then $A = (1+\beta)A_{1}/n$. At the end of Theorem proof we
set $\beta = 1/2$.

Denoting
$$
d_{i} = d(\mbox{\boldmath$x$}_{i}',\mbox{\boldmath $y$}') =
\|\mbox{\boldmath $y$}' - \mbox{\boldmath$x$}_{i}'\|^{2},
$$
arrange the distances $\{d_{i}, i= 1,\ldots,M\}$ for the
receiver after phase I in the increasing order, and denote
$$
d^{(1)} = \min_{i} d_{i} \leq  d^{(2)} \leq \ldots \leq d^{(M)} =
\max_{i} d_{i}
$$
(case of tie has zero probability). Let also
${\mbox{\boldmath $x$}'}^{(1)},\ldots,{\mbox{\boldmath $x$}'}^{(M)}$
be the corresponding ranking of codewords
$\{{\mbox{\boldmath $x$}'}\}$ after phase I for the receiver, i.e
${\mbox{\boldmath $x$}'}^{(1)}$ is the closest to
$\mbox{\boldmath $y$}'$ codeword, etc.

Similarly, denoting
$$
d_{i}^{(t)} = d(\mbox{\boldmath$x$}_{i}',\mbox{\boldmath $z$}') =
\|\mbox{\boldmath $z$}' - \mbox{\boldmath$x$}_{i}'\|^{2},
$$
arrange the distances $\{d_{i}^{(t)},\,i=1,\ldots,M\}$ for the
transmitter after phase I in the increasing order, denoting
$$
d^{(1)t} = \min_{i} d_{i}^{(t)}  \leq  d^{(2)t} \leq \ldots \leq
d^{(M)t} = \max_{i} d_{i}^{(t)}.
$$
Let also ${\mbox{\boldmath $x$}'}^{(1)t},\ldots,
{\mbox{\boldmath $x$}'}^{(M)t}$ be the corresponding ranking of
codewords $\{{\mbox{\boldmath $x$}'}\}$ after phase I for the
transmitter, i.e ${\mbox{\boldmath $x$}'}^{(1)t}$ is the closest to
$\mbox{\boldmath $z$}'$ codeword, etc.

{\bf Transmission method with one switching moment}. We choose a set
${\cal K}$ of codes ${\cal C}$ which the transmitter may use on
phase II. A code ${\cal C} \in {\cal K}$ used on phase II depends on
the received block $\mbox{\boldmath $z$}'$. Based on
$\mbox{\boldmath $y$}'$, the receiver finds the probability
distribution ${\mathbf P}_{r}({\cal C}|\mbox{\boldmath $y$}')$,
${\cal C} \in {\cal K}$ of the code ${\cal C}$ used by the
transmitter on phase II, and uses that distribution for optimal
decoding. It is a crucial point of the whole method.

{\bf Transmission}. In order to simplify exposition it is sufficient
to consider the case $M \leq (n+2)/2$. Then on both phases we will be
able to use orthogonal codes of length $n_{1} = n/2$. The case of
arbitrary $M$, such that $M = e^{o(n)}$, $n \to \infty$ can be
considered replacing orthogonal codes by ``almost'' equidistant
codes. Then all calculations remain essentially the same (see details
in \cite{BY2010}).

{\sl Phase I}. The transmitter uses the orthogonal code of $M$
codewords $\{\mbox{\boldmath $x$}_{i}'\}$ of length $n_{1}$
such that  $\|\mbox{\boldmath $x$}_{i}'\|^{2} = A_{1}$.

{\sl Phase II}. We set nonnegative numbers $\tau_{2}$ and $\tau_{3}$.
Based on the received block $\mbox{\boldmath $z$}'$ and numbers
$\tau_{2},\tau_{3}$, the transmitter chooses
$k = k(\mbox{\boldmath $z$}',\tau_{2}, \tau_{3})$ most probable
(for him) messages $k =2,3,4$. Denote that set of messages as
\begin{equation}\label{defSk}
{\cal S}^{k} = \left\{{\mbox{\boldmath $x$}'}^{(1)t},\ldots,
{\mbox{\boldmath $x$}'}^{(k)t}\right\}, \qquad k =2,3,4.
\end{equation}

The code length $n_{1}$ for phase II we partition on two parts:
of length $3$ for selected $k \in \{2,3,4\}$ messages and of length
$n_{1}-3$ for remaining $n - k$ messages, respectively. The
transmitter uses the following code
${\cal C}'' = {\cal C}''(\mbox{\boldmath $z$}')$ with
$\|\mbox{\boldmath $x$}_{j}''\|^{2} = A_{2}$, $j=1,\ldots,M$.

1) If $d^{(3)t} - d^{(2)t} \geq 2A_{1}\tau_{2}$, then the transmitter
selects two most probable (for him) messages $\theta_{i},\theta_{j}$
(i.e. $k = 2$) and uses for them opposite codewords
$\mbox{\boldmath $x$}_{i}'' = -\mbox{\boldmath $x$}_{j}''$
that have nonzero coordinates only at time instant $n_{1} + 1$.

For remaining $M-2$ messages $\{\theta_{s}\}$ the orthogonal code of
$M-2$ codewords $\{\mbox{\boldmath $x$}_{s}''\}$ of length
$n_{1}-3$ is used. That code have zero components at time instants
$n_{1}+1,n_{1}+2, n_{1}+3$, and all its codewords
$\{\mbox{\boldmath $x$}_{s}''\}$ are orthogonal to the first two
codewords $(\mbox{\boldmath $x$}_{i}'',\mbox{\boldmath $x$}_{j}'')$.

2) If $d^{(3)t} - d^{(2)t} < 2A_{1}\tau_{2}$,
$d^{(4)t} - d^{(3)t} \geq 2A_{1}\tau_{3}$
then the transmitter selects three most probable (for him) messages
$\theta_{i},\theta_{j},\theta_{m}$ (i.e. $k = 3$) and uses for them
the $3$-simplex code occupying time instants $n_{1}+1, n_{1}+2$.

For remaining $M-3$ messages $\{\theta_{s}\}$ the orthogonal code of
codewords $\{\mbox{\boldmath $x$}_{s}''\}$ of length $n_{1}-3$ is
used. That code have zero components at time instants
$n_{1}+1,n_{1}+2, n_{1}+3$, and all its codewords
$\{\mbox{\boldmath $x$}_{s}''\}$ are orthogonal to the first three
codewords $(\mbox{\boldmath $x$}_{i}'',\mbox{\boldmath $x$}_{j}'',
\mbox{\boldmath $x$}_{m}'')$.

3) If $d^{(3)t} - d^{(2)t} < 2A_{1}\tau_{2}$,
$d^{(4)t} - d^{(3)t} < 2A_{1}\tau_{3}$, then the transmitter selects
four most probable (for him) messages
$\theta_{i},\theta_{j},\theta_{m},\theta_{l}$ (i.e. $k = 4$)
and uses for them the $4$-simplex code, occupying time instants
$n_{1}+1, n_{1}+2,n_{1}+3$.

For remaining $M-4$ messages $\{\theta_{s}\}$ the orthogonal code of
codewords $\{\mbox{\boldmath $x$}_{s}''\}$ of length $n_{1}-3$ is
used. That code have zero components at time instants
$n_{1}+1,n_{1}+2, n_{1}+3$, and all its codewords
$\{\mbox{\boldmath $x$}_{s}''\}$ are orthogonal to the first four
codewords $(\mbox{\boldmath $x$}_{i}'',\mbox{\boldmath $x$}_{j}'',
\mbox{\boldmath $x$}_{m}'',\mbox{\boldmath $x$}_{l}'')$.

This transmission method strengthens the method used in
\cite{BY2010}, \cite{BY0}--\cite{BY2}, where only
two or three messages were selected.

Note also that the set ${\cal S}^{k}$ of selected messages should be
such that with high probability the true message
$\theta_{\rm true} \in {\cal S}^{k}$, but the number $k$ is small as
possible.

{\it Remark} 3.
Introducing additional parameters $\tau_4,\ldots\strut$,  it is
possible to strengthen the method used, but it gives not a big
improvement of the results obtained. Much more improvement can be
obtained using an increasing number of $N=N(\sigma)$ (see remark 2).

{\bf Decoding}. Due to noise in the feedback channel the receiver
does not know exactly codewords
${\mbox{\boldmath $x$}'}^{(1)t},{\mbox{\boldmath $x$}'}^{(2)t},
\ldots$
and therefore it does not know the code used on phase II. But based
on the received block $\mbox{\boldmath $y$}'$ it may evaluate
probabilities of all possible codewords
${\mbox{\boldmath $x$}'}^{(1)t},{\mbox{\boldmath $x$}'}^{(2)t},
\ldots$
and find the probabilities with which any code ${\cal C}''$
was used on phase II.
It allows to the receiver, based on the full received block
$\mbox{\boldmath $y$}=(\mbox{\boldmath $y$}',\mbox{\boldmath $y$}'')$,
to find posterior probabilities
$\{p(\mbox{\boldmath $y$}|\mbox{\boldmath $x$}_{i})\}$
and make decision in favor of most probable message $\theta_{i}$.
Such full decoding is described in details in the next section.

\begin{center}
{\large\bf \S\,3. Full decoding and error probability $P_{\rm e}$}
\end{center}

Since $\|\mbox{\boldmath $x$}_{i}\|^{2} = A$, $i=1,\ldots,M$,
for the likelihood ratio we have
$$
\ln\frac{p\left(\mbox{\boldmath $y$}|\mbox{\boldmath $x$}_{i}\right)}
{p\left(\mbox{\boldmath $y$}|\mbox{\boldmath $x$}_{1}\right)} =
(\mbox{\boldmath $x$}_{i} - \mbox{\boldmath $x$}_{1},
\mbox{\boldmath $y$}).
$$
If $\mbox{\boldmath $x$}_{\rm true}$ is the true codeword then
$\mbox{\boldmath $y$} = \mbox{\boldmath $x$}_{\rm true} +
\mbox{\boldmath $\xi$}$ and
$\mbox{\boldmath $\xi$} = (\mbox{\boldmath $\xi$}',
\mbox{\boldmath $\xi$}'') = (\xi_{1},\ldots,\xi_{n})$, where
all $\{\xi_{i}\}$ are independent ${\cal N}(0,1)$--Gaussian random
variables. If
$\mbox{\boldmath $x$}_{\rm true} =  \mbox{\boldmath $x$}_{1}$, then
$$
\ln\frac{p\left(\mbox{\boldmath $y$}|
\mbox{\boldmath $x$}_{i}\right)}
{p\left(\mbox{\boldmath $y$}|\mbox{\boldmath $x$}_{1}\right)} =
(\mbox{\boldmath $x$}_{i} - \mbox{\boldmath $x$}_{1},
\mbox{\boldmath $\xi$}) - \frac{1}{2}
\|\mbox{\boldmath $x$}_{i} - \mbox{\boldmath $x$}_{1}\|^{2}
$$
and
$$
\ln\frac{p\left(\mbox{\boldmath $y$}|
\mbox{\boldmath $x$}_{3}\right)}
{p\left(\mbox{\boldmath $y$}|\mbox{\boldmath $x$}_{2}\right)} =
(\mbox{\boldmath $x$}_{3} - \mbox{\boldmath $x$}_{2},
\mbox{\boldmath $\xi$}) +
(\mbox{\boldmath $x$}_{3} - \mbox{\boldmath $x$}_{2},
\mbox{\boldmath $x$}_{1}),
$$
where $(\mbox{\boldmath $x$},\mbox{\boldmath $\xi$})$ is
${\cal N}(0,\|\mbox{\boldmath $x$}\|^{2})$--Gaussian random variable.

For decoding error probability $P_{\rm e}$ we have
\begin{equation}\label{genPe}
\begin{gathered}
P_{\rm e} \leq \frac{1}{M}\sum\limits_{k =1}^{M}P_{{\rm e}k},
\end{gathered}
\end{equation}
where
\begin{equation}\label{genPek}
\begin{gathered}
P_{{\rm e}k} = {\mathbf P}\left\{\max_{i \neq k}
\ln \frac{p\left({\mbox{\boldmath $y$}}\big|\theta_{i}\right)}
{p\left({\mbox{\boldmath $y$}}\big|\theta_{k}\right)} \geq 0
\big|\theta_{k}\right\}, \quad k =1,\ldots,M.
\end{gathered}
\end{equation}
Denote ($(\mbox{\boldmath $x$}_{i}',\mbox{\boldmath $x$}_{1}') = 0$,
$i \geq 2$)
\begin{equation}\label{XY}
\begin{gathered}
X_{i}= \ln\frac{p\left({\mbox{\boldmath $y$}'}\big|\theta_{i}\right)}
{p\left({\mbox{\boldmath $y$}'}\big|\theta_{1}\right)} =
(\mbox{\boldmath $x$}_{i}' -\mbox{\boldmath $x$}_{1}',
\mbox{\boldmath $y$}') = (\mbox{\boldmath $x$}_{i}' -
\mbox{\boldmath $x$}_{1}',\mbox{\boldmath $\xi$}') - A_{1}, \\
Y_{i} = \ln \frac{p\left({\mbox{\boldmath $y$}''}\big|
{\mbox{\boldmath $y$}'},\theta_{i}\right)}
{p\left({\mbox{\boldmath $y$}''}\big|{\mbox{\boldmath $y$}'},
\theta_{1}\right)}.
\end{gathered}
\end{equation}
It is sufficient to investigate the value $P_{{\rm e}1}$, for which
we have from \eqref{genPek}--\eqref{XY}
\begin{equation}\label{stransPe1}
\begin{gathered}
P_{{\rm e}1} = {\mathbf P}
\left\{\max_{i \geq 2}(X_{i} + Y_{i}) \geq 0
\big|\theta_{1}\right\} \leq \sum\limits_{i \geq 2}
{\mathbf P}\left\{X_{i} + Y_{i} \geq 0\big|\theta_{1}\right\} = \\
= \sum\limits_{i \geq 2}{\mathbf E}_{\mbox{\boldmath $y$}'}
{\mathbf P}\left\{X_{i} + Y_{i} \geq 0
\big|\mbox{\boldmath $y$}',\theta_{1}\right\}.
\end{gathered}
\end{equation}
We can express the value $Y_{i}$ via $\mbox{\boldmath $y$}'$ as
follows. Since
$\mbox{\boldmath $x$}_{i}'' = \mbox{\boldmath $x$}_{i}''
(\mbox{\boldmath $z$}')$ and $\mbox{\boldmath $y$}'' =
\mbox{\boldmath $x$}_{1}'' + \mbox{\boldmath $\xi$}''$, then
\begin{equation}\label{defeY}
\begin{gathered}
e^{Y_{i}} = \frac{p\left({\mbox{\boldmath $y$}''}\big|
{\mbox{\boldmath $y$}'},\theta_{i}\right)}
{p\left({\mbox{\boldmath $y$}''}\big|{\mbox{\boldmath $y$}'},
\theta_{1}\right)} =
{\mathbf E}_{\mbox{\boldmath $z$}'|\mbox{\boldmath $y$}'}
\frac{p\left(\mbox{\boldmath $y$}''\big|\mbox{\boldmath $z$}',
\mbox{\boldmath $y$}',\mbox{\boldmath $x$}_{i}''\right)}
{p\left(\mbox{\boldmath $y$}''
\big|\mbox{\boldmath $z$}',\mbox{\boldmath $y$}',
\mbox{\boldmath $x$}_{1}''\right)} = \\
= {\mathbf E}_{\mbox{\boldmath $z$}'|\mbox{\boldmath $y$}'}
e^{(\mbox{\boldmath $y$}'',\mbox{\boldmath $x$}_{i}'' -
\mbox{\boldmath $x$}_{1}'')} =
{\mathbf E}_{\mbox{\boldmath $z$}'|\mbox{\boldmath $y$}'}
e^{(\mbox{\boldmath $x$}_{1}'',\mbox{\boldmath $x$}_{i}'' -
\mbox{\boldmath $x$}_{1}'') +
(\mbox{\boldmath $\xi$}'',\mbox{\boldmath $x$}_{i}'' -
\mbox{\boldmath $x$}_{1}'')},
\end{gathered}
\end{equation}
where the second equality is based on the fact that in both
nominator and denominator the same code is used.

{\it Remark} 4. In order to apply the formula \eqref{defeY} it is
necessary to know only the difference
$\|\mbox{\boldmath $x$}_{i}'' -\mbox{\boldmath $x$}_{1}''\|$
(depending on $\mbox{\boldmath $z$}'$). We do not need to know the
whole code used on phase II. The selected group of messages of the
code for phase II may consist of $2,3,4$ messages. For example, $3$
messages are selected if $3$ most probable messages are
approximately equiprobable and all remaining messages are well
separated from them (in metrics $d_{i}^{(t)}$).

We develop the right-hand side of the formula \eqref{defeY}. The
difference
$\|\mbox{\boldmath $x$}_{i}'' -\mbox{\boldmath $x$}_{1}''\|$
(depending on $\mbox{\boldmath $z$}'$) takes on one of $4$ possible
values (defined by partition groups, which those messages belong to
on phase II). It is convenient to separate those cases. Note that
for all codewords of the $k$-simplex code we have
$$
\begin{gathered}
d_{ij}=\|\mbox{\boldmath $x$}_{i}'' -\mbox{\boldmath $x$}_{j}''\|^{2}
= 2A_{2}k/(k-1), \qquad i \neq j.
\end{gathered}
$$
Then denote
\begin{equation}\label{simp1}
\begin{gathered}
\delta_{k} = 2A_{2}k/(k-1), \qquad k = 2,\ldots,K, \\
\delta_{0} = 2A_{2}.
\end{gathered}
\end{equation}
In other words, $\delta_{k}$, $k \geq 2$ is the codewords distance
for $k$-simplex code, while $d_{0}$ is such distance for the
orthogonal code. If
$\mbox{\boldmath $x$}_{1}'',\mbox{\boldmath $x$}_{i}''$
belong to $k$-simplex code then
$$
\begin{gathered}
(\mbox{\boldmath $x$}_{1}'',\mbox{\boldmath $x$}_{i}'' -
\mbox{\boldmath $x$}_{1}'') = -\delta_{k}/2 = -A_{2}k/(k-1),
\qquad k = 2,\ldots,K, \\
(\mbox{\boldmath $x$}_{1}'',\mbox{\boldmath $x$}_{i}'' -
\mbox{\boldmath $x$}_{1}'') = -A_{2}= -\delta_{0}/2, \qquad k = 0.
\end{gathered}
$$

The difference
$\|\mbox{\boldmath $x$}_{i}'' -\mbox{\boldmath $x$}_{1}''\|$ may
take on values $2A_{2}$ (corresponds to $k=0$) and
$\delta_{k}$, $k = 2,3,4$. Each value $\delta_{k}$, $k = 2,3,4$
appears for phase II if a group of $k$ messages was selected and
both messages
$\mbox{\boldmath $x$}_{1}',\mbox{\boldmath $x$}_{i}'$ belong to that
group. In all other cases the value $\delta_{0}$ is used.

Assuming $\theta_{\rm true} = \theta_{1}$, introduce non-overlapping
sets  of random events
\begin{equation}\label{defZi2}
\begin{gathered}
{\cal Z}_{i,k} = \left\{\mbox{\boldmath $z$}':
\|\mbox{\boldmath $x$}_{i}'' -\mbox{\boldmath $x$}_{1}''\|^{2} =
\delta_{k}\right\}, \qquad k = 0,2,3,4.
\end{gathered}
\end{equation}
Denoting formally ${\cal Z}_{i,1} = \emptyset$, $i \geq 2$, we have
$\{\mbox{\boldmath $z$}'\} = \sum\limits_{k=0}^{4}{\cal Z}_{i,k}$
(here $\sum$ means the union of non-overlapping sets, and
\{\mbox{\boldmath $z'$}\} is the set of all possible outputs
$\mbox{\boldmath $z$}'$).

We may continue \eqref{defeY} as follows
$$
\begin{gathered}
e^{Y_{i}} = \sum\limits_{k=0}^{4}
{\mathbf E}\left[e^{(\mbox{\boldmath $x$}_{1}'',
\mbox{\boldmath $x$}_{i}'' -\mbox{\boldmath $x$}_{1}'') +
(\mbox{\boldmath $\xi$}'',\mbox{\boldmath $x$}_{i}'' -
\mbox{\boldmath $x$}_{1}'')};{\cal Z}_{i,k}\Big|\mbox{\boldmath $y$}'
\right] = \sum\limits_{k=0}^{4}p_{k}e^{-\delta_{k}/2 +
(\mbox{\boldmath $\xi$}'',\mbox{\boldmath $x$}_{i}'' -
\mbox{\boldmath $x$}_{1}'')},
\end{gathered}
$$
where ${\mathbf E}[\xi;{\cal A}] =
{\mathbf E}(\xi \cdot I_{\{{\cal A}\}})$, $p_{1} = 0$ and
\begin{equation}\label{defpk}
\begin{gathered}
p_{k} = p_{k}(\mbox{\boldmath $y$}') =
p_{k}(\mbox{\boldmath $\xi$}') =
{\mathbf P}\left({\cal Z}_{i,k}\big|\mbox{\boldmath $y$}'\right),
\qquad k = 0,2,3,4.
\end{gathered}
\end{equation}
Then using \eqref {XY} we have
$$
\begin{gathered}
e^{X_{i} + Y_{i}} = \sum\limits_{k=0}^{4}p_{k}e^{-A_{1}-\delta_{k}/2
+ (\mbox{\boldmath $x$}_{i}' -\mbox{\boldmath $x$}_{1}',
\mbox{\boldmath $\xi$}') + (\mbox{\boldmath $\xi$}'',
\mbox{\boldmath $x$}_{i}'' - \mbox{\boldmath $x$}_{1}'')},
\end{gathered}
$$
and therefore (since $k$ takes on one of four possible values)
\begin{equation}\label{defpk1}
\begin{gathered}
{\mathbf P}\left\{X_{i} + Y_{i} \geq 0\big|\theta_{1}\right\} =
{\mathbf E}{\mathbf P}\left\{e^{X_{i} + Y_{i}} \geq 1
\big|\mbox{\boldmath $y$}',\theta_{1}\right\} = \\
= {\mathbf E}{\mathbf P}\left\{\sum\limits_{k=0}^{4}p_{k}
e^{-A_{1}-\delta_{k}/2 +
(\mbox{\boldmath $x$}_{i}' - \mbox{\boldmath $x$}_{1}',
\mbox{\boldmath $\xi$}') + (\mbox{\boldmath $x$}_{i}'' -
\mbox{\boldmath $x$}_{1}'',\mbox{\boldmath $\xi$}'')} \geq 1
\big|\mbox{\boldmath $y$}',\theta_{1}\right\} \leq \\
\leq \sum\limits_{k=0}^{4}{\mathbf E}{\mathbf P}
\left\{\left[p_{k}e^{-A_{1}-\delta_{k}/2 +
(\mbox{\boldmath $x$}_{i}' - \mbox{\boldmath $x$}_{1}',
\mbox{\boldmath $\xi$}') + (\mbox{\boldmath $x$}_{i}'' -
\mbox{\boldmath $x$}_{1}'',\mbox{\boldmath $\xi$}'')} \geq 1/4\right]
\bigcap {\cal Z}_{i,k}\big|\mbox{\boldmath $y$}',\theta_{1}\right\} =
\\ = \sum\limits_{k=0}^{4}{\mathbf P}\left\{\left[
(\mbox{\boldmath $x$}_{i}' - \mbox{\boldmath $x$}_{1}',
\mbox{\boldmath $\xi$}') + (\mbox{\boldmath $x$}_{i}'' -
\mbox{\boldmath $x$}_{1}'',\mbox{\boldmath $\xi$}'') +
\ln p_{k}(\mbox{\boldmath $\xi$}') \geq
A_{1}+\delta_{k}/2 - \ln 4\right]\bigcap {\cal Z}_{i,k}\big|
\theta_{1}\right\},
\end{gathered}
\end{equation}
where $\|\mbox{\boldmath $x$}_{i}'' -\mbox{\boldmath $x$}_{1}''\|^{2}
= \delta_{k}$ for the set ${\cal Z}_{i,k}$. Denote
\begin{equation}\label{defxieta1}
\begin{gathered}
(\mbox{\boldmath$x$}_{i}',\mbox{\boldmath $\xi$}') =
\sqrt{A_{1}}\xi_{i}', \qquad
(\mbox{\boldmath$x$}_{i}',\mbox{\boldmath $\eta$}') =
\sqrt{A_{1}}\eta_{i}', \qquad  i = 1,\ldots,M,
\end{gathered}
\end{equation}
where all $\{\xi_{i}',\eta_{i}'\}$ are independent
${\cal N}(0,1)$-Gaussian random variables.

Since $(\mbox{\boldmath $x$}_{i}'' - \mbox{\boldmath $x$}_{1}'',
\mbox{\boldmath $\xi$}'') \sim \sqrt{\delta_{k}}\xi''$ for the set
${\cal Z}_{i,k}$, we get from \eqref{defpk1} and \eqref{defxieta1}
\begin{equation}\label{defpk1a}
\begin{gathered}
{\mathbf P}\left\{X_{i} + Y_{i} \geq 0 \big|\theta_{1}\right\} \leq
e^{o(1)}\sum\limits_{k=0}^{4}P_{ik}, \\
P_{ik} = {\mathbf P}\left\{\sqrt{A_{1}}(\xi_{i}' - \xi_{1}') +
\sqrt{d_{k}}\xi'' + \ln p_{k}(\mbox{\boldmath $\xi$}')
\geq A_{1}+ \delta_{k}/2 \right\},
\end{gathered}
\end{equation}
where $\xi''$ does not depend on $\mbox{\boldmath $\xi$}'$
and $o(1) \to 0$ as $A_{1} \to \infty$.

Probabilities $\{p_{k}(\mbox{\boldmath $\xi$}')\}$ and values
$P_{ik}$ from \eqref{defpk1a} are evaluated in the next section.

\begin{center}
{\large\bf \S\,4. Probabilities $p_{k}(\mbox{\boldmath $\xi$}')$
and values  $P_{ik}$.  Proof of Theorem}
\end{center}

Let $\xi$ be ${\cal N}(0,1)$--Gaussian random variable.
We will regularly use simple inequality
\begin{equation}\label{Phi1}
{\mathbf P}(\xi \geq z) = \frac{1}{\sqrt{2\pi}}\int\limits_{z}^{\infty}
e^{-u^{2}/2}du \leq e^{-z_{+}^{2}/2}, \qquad z \in \mathbb{R}^{1},
\end{equation}
and its natural generalization

L e m m a \,1. 1) {\it Let $(\xi_{1},\ldots,\xi_{K})$ be independent
${\cal N}(0,1)$--Gaussian random variables and
${\cal A} \subseteq \mathbb{R}^{K}$. Then}
($\mbox{\boldmath $x$} = (x_{1},\ldots,x_{K})$,
$\|\mbox{\boldmath $x$}\|^{2} = x_{1}^{2} + \ldots + x_{K}^{2}$)
\begin{equation}\label{Phi2}
\begin{gathered}
{\mathbf P}\left((\xi_{1},\ldots,\xi_{K}) \in {\cal A}\right) \leq
\exp\left\{-\frac{1}{2}\inf_{\mbox{\boldmath $x$} \in {\cal A}}
\|\mbox{\boldmath $x$}\|^{2}\right\}.
\end{gathered}
\end{equation}

2) {\it Let $\xi,\eta$ be ${\cal N}(0,1)$--Gaussian random variables
and ${\mathbf E}(\xi \eta) = \rho$. Then}:

a) {\it if} $A-B\rho \geq 0$ and $B-A\rho \geq 0$ then
\begin{equation}\label{lem1}
\begin{gathered}
{\mathbf P}\left(\xi \geq A, \eta \geq B\right) \leq {\mathbf P}
\left(\xi \geq \sqrt{\frac{A^{2}+B^{2} -2AB\rho}{1-\rho^{2}}}\right);
\end{gathered}
\end{equation}

b) {\it otherwise}
\begin{equation}\label{lem1a}
\begin{gathered}
{\mathbf P}\left(\xi \geq A, \eta \geq B\right) \leq
\min\left\{{\mathbf P}(\xi \geq A), {\mathbf P}(\eta \geq B)\right\}.
\end{gathered}
\end{equation}

P r o o f. 1) Let $\inf\limits_{\mbox{\boldmath $x$} \in
{\cal A}}\|\mbox{\boldmath $x$}\| = r > 0$. Then
${\cal A} \subseteq \mathbb{R}^{K} \setminus S(r)$, where
$S(r)$ -- the ball of radius $r$. Therefore
$$
\begin{gathered}
{\mathbf P}\left((\xi_{1},\ldots,\xi_{K}) \in {\cal A}\right) \leq
{\mathbf P}\left\{(\xi_{1},\ldots,\xi_{K}) \in \mathbb{R}^{K} \setminus
S(r)\right\}.
\end{gathered}
$$
Evaluating the last probability (using spherical coordinates) we get
the formula \eqref{Phi2}.

2) We have
$$
\begin{gathered}
{\mathbf P}(\xi \geq A, \eta \geq B) \leq \inf_{a \geq 0}{\mathbf P}
(\xi + a\eta \geq A + aB) = {\mathbf P}
\left(\xi \geq \frac{A + aB}{\sqrt{1+a^{2} + 2a\rho}}\right).
\end{gathered}
$$
Minimizing the last expression over $a \geq 0$, we get
the formulas \eqref{lem1}--\eqref{lem1a}.  \qquad $\Box$

Inequalities \eqref{Phi2}--\eqref{lem1a} give the exact logarithmic
asymptotics in a natural asymptotic case.

In order to apply the formula \eqref{defpk1a}, we consider
sequentially the cases $k = 2,0,3,4$.

{\bf 1. Case $\mathbf{k = 2}$}, $\delta_{2} = 4A_{2}$. It is the
simplest case and it takes place with probability close to $1$.
In that case $\mbox{\boldmath $x$}_{1}'',\mbox{\boldmath $x$}_{i}''$
compose the group ${\cal S}^{2}$ of two selected messages. Neglecting
the term $p_{2}$ we get from \eqref{defpk1a}--\eqref{Phi1}
\begin{equation}\label{EPi2}
\begin{gathered}
P_{i2} \leq {\mathbf P}\left\{
(\mbox{\boldmath $x$}_{i}' -\mbox{\boldmath $x$}_{1}',
\mbox{\boldmath $\xi$}') + 2\sqrt{A_{2}}\xi''
\geq A_{1}+ 2A_{2} - \ln 3\right\} = \\
= {\mathbf P}\left\{\sqrt{2A_{1} + 4A_{2}}\xi
\geq A_{1}+ 2A_{2} - \ln 3\right\} \leq
\exp\left\{- \frac{[A_{1}+ 2A_{2} - \ln 3]_{+}^{2}}{4(A_{1}+ 2A_{2})}
\right\} \leq \\
\leq \sqrt{3}e^{-(A_{1}+2A_{2})/4} = \sqrt{3}e^{-A_{1}(1+2\beta)/4}.
\end{gathered}
\end{equation}

Cases $k \neq 2$ are more computationally involved and in order to
investigate them we will need the definition \eqref{defSk}.

{\bf 2. Case $\mathbf{k = 0}$}, $\delta_{0} = 2A_{2}$. It is the most
computationally involved case. It takes place when the selected
group of messages ${\cal S}^{m}$ contains  not more than one of
messages $\mbox{\boldmath $x$}_{1}'',\mbox{\boldmath $x$}_{i}''$.
Then
\begin{equation}\label{Pi0}
\begin{gathered}
P_{i0} = \sum\limits_{m=2}^{4}P_{i0m},
\end{gathered}
\end{equation}
where $P_{i0m} = P\{k = 0,{\cal S}^{m}\}$, $m = 2,3,4$. We consider
sequentially probabilities $\{P_{i0m}, m = 2,3,4\}$,
starting with $P_{i02}$. Denote
$$
\begin{gathered}
d'_{ij}= \|\mbox{\boldmath$x$}_{i}' - \mbox{\boldmath$x$}_{j}'\|^{2}.
\end{gathered}
$$
If $\theta_{\rm true} = \theta_{1}$ then the formulas hold
\begin{equation}\label{gendd}
\begin{gathered}
d_{i} - d_{j} = d'_{1i} - d'_{1j} +
2(\mbox{\boldmath$x$}_{j}' - \mbox{\boldmath$x$}_{i}',
\mbox{\boldmath $\xi$}'),  \qquad i,j = 1,\ldots,M, \\
d_{i} - d_{1} = d'_{1i} + 2(\mbox{\boldmath$x$}_{1}' -
\mbox{\boldmath$x$}_{i}',\mbox{\boldmath $\xi$}'), \\
d_{i}^{(t)} - d_{j}^{(t)} = d'_{1i} - d'_{1j} +
2(\mbox{\boldmath$x$}_{j}' - \mbox{\boldmath$x$}_{i}',
\mbox{\boldmath $\xi$}' + \sigma \mbox{\boldmath $\eta$}') =
d_{i} - d_{j} + 2\sigma (\mbox{\boldmath$x$}_{j}' -
\mbox{\boldmath$x$}_{i}',\mbox{\boldmath $\eta$}'), \\
d_{i}^{(t)} - d_{1}^{(t)} = d'_{1i} +
2\left(\mbox{\boldmath $x$}_{1}' - \mbox{\boldmath$x$}_{i}',
\mbox{\boldmath $\xi$}' + \sigma \mbox{\boldmath $\eta$}'\right) =
d_{i} - d_{1} + 2\sigma (\mbox{\boldmath$x$}_{1}' -
\mbox{\boldmath$x$}_{i}',\mbox{\boldmath $\eta$}').
\end{gathered}
\end{equation}
If, in particular,
${\mbox{\boldmath $x$}'}^{(1)t} = \mbox{\boldmath $x$}_{1}'$,
${\mbox{\boldmath $x$}'}^{(2)t} = \mbox{\boldmath $x$}_{2}'$, and
${\mbox{\boldmath $x$}'}^{(3)t} = {\mbox{\boldmath $x$}_{i}'}$,
$i \geq 3$, then in the case ${\cal S}^{2}$ it is necessary to have
\begin{equation}\label{S2}
\begin{gathered}
d_{3}^{(t)} - d_{2}^{(t)} =
2(\mbox{\boldmath$x$}_{2}' - \mbox{\boldmath$x$}_{i}',
\mbox{\boldmath $\xi$}' + \sigma \mbox{\boldmath $\eta$}')
\geq 2A_{1}\tau_{2}.
\end{gathered}
\end{equation}

In order to evaluate $p_{0} = p_{0}(\mbox{\boldmath $\xi$}')$ from
\eqref{defpk} notice that the main contribution to $p_{0}$ gives
the case when the true message $\mbox{\boldmath $x$}_{1}'$ is
selected, but the message $\mbox{\boldmath $x$}_{i}'$ is not.
Moreover, in the case ${\cal S}^{2}$ maximum of $p_{0}$ is attained
when ${\mbox{\boldmath $x$}'}^{(1)t} = {\mbox{\boldmath $x$}_{1}'}$,
${\mbox{\boldmath $x$}_{i}'} \not\in
\{{\mbox{\boldmath $x$}'}^{(1)t},{\mbox{\boldmath $x$}'}^{(2)t}\}$.
Taking into account symmetry of the orthogonal code
$\{{\mbox{\boldmath $x$}_{j}'}\}$, we may assume that
${\mbox{\boldmath $x$}'}^{(2)t} = {\mbox{\boldmath $x$}_{2}'}$ and
${\mbox{\boldmath $x$}'}^{(3)t} = {\mbox{\boldmath $x$}_{i}'}$,
$i \geq 3$. Since there are not more than $M^{3}$ variants of
arranging messages
${\mbox{\boldmath $x$}_{1}'},{\mbox{\boldmath $x$}_{2}'},
{\mbox{\boldmath $x$}_{i}'}$, then using \eqref{S2}, we have
\begin{equation}\label{S2a}
\begin{gathered}
p_{0}(\mbox{\boldmath $\xi$}') \leq M^{3}{\mathbf P}\left(
(\mbox{\boldmath$x$}_{2}' - \mbox{\boldmath$x$}_{i}',
\mbox{\boldmath $\xi$}' + \sigma \mbox{\boldmath $\eta$}')
\geq A_{1}\tau_{2}\Big|\{\xi_{i}'\}\right) \leq \\
\leq M^{3}{\mathbf P}\left(
\sigma(\mbox{\boldmath$x$}_{2}' - \mbox{\boldmath$x$}_{i}',
\mbox{\boldmath $\eta$}') \geq A_{1}\tau_{2} + \sqrt{A_{1}}\xi_{i}' -
\sqrt{A_{1}}\xi_{2}'\Big|\{\xi_{i}'\}\right) \leq \\
\leq M^{3}\exp\left\{-\frac{
(\sqrt{A_{1}}\tau_{2}-\xi_{2}' + \xi_{i}')_{+}^{2}}
{4\sigma^{2}}\right\}.
\end{gathered}
\end{equation}
Since $M = e^{o(A_{1})}$, $A_{1} \to \infty$
(see \eqref{A1A2II}--\eqref{defbeta}), then from \eqref{defpk1a} and
\eqref{S2a} for $P_{i02}$ we get
\begin{equation}\label{S2b}
\begin{gathered}
P_{i02} \leq e^{o(A_{1})}{\mathbf P}\left\{\sqrt{2A_{2}}\xi'' +
\sqrt{A_{1}}(\xi_{i}' - \xi_{1}') -
\frac{(\sqrt{A_{1}}\tau_{2}+ \xi_{i}'-\xi_{2}')_{+}^{2}}
{4\sigma^{2}} \geq A_{1}+A_{2}\right\} = \\
= e^{o(A_{1})}{\mathbf P}
\left\{\sqrt{A_{1}}\left(\sqrt{3+4\beta}\zeta + \zeta_{2}\right) -
\frac{(\sqrt{A_{1}/2}\,\tau_{2}+ \zeta_{2})_{+}^{2}}
{\sigma^{2}\sqrt{2}} \geq A_{1}(1+\beta)\sqrt{2} \right\},
\end{gathered}
\end{equation}
where we used the representations
$\xi_{i}'-\xi_{2}' = \sqrt{2}\zeta_{2}$,
$\xi_{i}'-\xi_{1}' = \zeta_{2}/\sqrt{2} + \sqrt{3/2}\,\xi$,
$\xi \bot \zeta_{2}$ and $\sqrt{2A_{2}}\xi'' + \sqrt{3A_{1}/2}\,\xi =
\sqrt{(3+4\beta)A_{1}/2}\zeta$, $\zeta \bot \zeta_{2}$
(here $\xi \bot \zeta$ means that Gaussian random variables
$\xi,\zeta$ are orthogonal, i.e. independent).

Denoting $x\sqrt{A_{1}} = \zeta$, $y\sqrt{A_{1}} = \zeta_{2}$ and
using the formula \eqref{Phi2}, we have from \eqref{S2b}
\begin{equation}\label{Pi01}
\begin{gathered}
-2\ln P_{i02} \geq A_{1}\inf_{(x,y) \in {\cal A}}
\left(x^{2} + y^{2}\right) + o(A_{1}),  \\
{\cal A} = \left\{x,y: \sqrt{3+4\beta}x + y -
\frac{(\tau_{2}/\sqrt{2} + y)_{+}^{2}}
{\sigma^{2}\sqrt{2}} \geq (1+\beta)\sqrt{2}\right\}.
\end{gathered}
\end{equation}

Denoting
${\cal A}_{1} = \left\{y: y \leq -\tau_{2}/\sqrt{2}\right\}$,
first we have
$$
\begin{gathered}
\inf_{(x,y) \in ({\cal A}\cap {\cal A}_{1})}(x^{2} + y^{2}) =
\inf_{\substack{\sqrt{3+4\beta}x + y \geq (1+\beta)\sqrt{2} \\
y \leq -\tau_{2}/\sqrt{2}}}(x^{2} + y^{2})  =
\inf_{\substack{\sqrt{3+4\beta}x + y \geq (1+\beta)\sqrt{2} \\
y = -\tau_{2}/\sqrt{2}}}(x^{2} + y^{2}),
\end{gathered}
$$
i.e. infimum is attained on the border of ${\cal A}$.
Therefore we may assume that $\tau_{2}/\sqrt{2} + y \geq 0$, omit
the sign of positive part and replace \eqref{Pi01} by
\begin{equation}\label{Pi02}
\begin{gathered}
-2\ln P_{i02} \geq A_{1}\inf_{(x,y) \in {\cal A}_{2}}
\left(x^{2} + y^{2}\right) + o(A_{1}),  \\
{\cal A}_{2} =\left\{x,y: x -\varepsilon (y + a)^{2} \geq B \right\},
\end{gathered}
\end{equation}
where
$$
\begin{gathered}
\varepsilon = \frac{1}{\sigma^{2}\sqrt{2(3+4\beta)}}, \qquad
a = \frac{\sqrt{2}(\tau_{2}-\sigma^{2})}{2}, \qquad
B = \sqrt{\frac{2}{3+4\beta}}\left(1+\beta +
\frac{2\tau_{2} - \sigma^{2}}{4}\right).
\end{gathered}
$$
If $B \geq 0$, then for optimal $x,y$ we need
$x - \varepsilon (y + a)^{2} = B$ and therefore (omitting
$\varepsilon^{2}(y +a)^{4}$)
$$
\begin{gathered}
\inf_{(x,y) \in {\cal A}_{2}}\left(x^{2} + y^{2}\right) =
\inf_{y}\left\{[\varepsilon (y +a)^{2} +B]^{2} +y^{2}\right\} \geq \\
\geq \inf_{y}\left\{2B\varepsilon (y +a)^{2} +B^{2}+y^{2}\right\}
= B^{2} + a^{2} - \frac{a^{2}}{1+2B\varepsilon} \geq
B^{2} + a^{2} - \frac{a^{2}}{2B\varepsilon}.
\end{gathered}
$$
Therefore we get from \eqref{Pi02} as $A_{1} \to \infty$
\begin{equation}\label{Pi03}
\begin{gathered}
-\ln P_{i02} \geq \frac{A_{1}}{2}\left\{
\frac{(4 + 4\beta + 2\tau_{2} - \sigma^{2})^{2}}{8(3+4\beta)} +
\frac{(\tau_{2}-\sigma^{2})^{2}}{2} -
\frac{(\tau_{2}-\sigma^{2})^{2}(3+4\beta)\sigma^{2}}
{4 + 4\beta + 2\tau_{2} - \sigma^{2}} + o(1)\right\}.
\end{gathered}
\end{equation}
We consider below only $\sigma^{2} \leq 1$ and $\tau_{2} \leq 4/9$.
Then we can simplify the formula \eqref{Pi03} as follows
\begin{equation}\label{Pi02sim}
\begin{gathered}
-\ln P_{i02} \geq \frac{A_{1}}{2}
\left\{\frac{(2 + 2\beta + \tau_{2})^{2}}{2(3+4\beta)} +
\frac{\tau_{2}^{2}}{2} - \sigma^{2}\left[\frac{2 + 2\beta + \tau_{2}}
{2(3+4\beta)} + \tau_{2} + \tau_{2}^{2}\right] + o(1)\right\} \geq \\
\geq \frac{A_{1}}{4}\left[\frac{(2 + 2\beta + \tau_{2})^{2}}
{3+4\beta} + \tau_{2}^{2} + o(1)\right](1-\sigma^{2}) = \\
= \frac{A_{1}(1+\beta)(1+\beta + \tau_{2} + \tau_{2}^{2})
(1-\sigma^{2})}{3+4\beta} + o(A_{1}),
\end{gathered}
\end{equation}
since
$$
\frac{(2 + 2\beta + \tau_{2})^{2}}{2(3+4\beta)} +
\frac{\tau_{2}^{2}}{2} \geq \frac{2 + 2\beta + \tau_{2}}
{2(3+4\beta)} + \tau_{2} + \tau_{2}^{2}, \qquad \tau_{2} \leq 4/9.
$$

Consider the case of ${\cal S}^{3}$ and $P_{i03}$. Again, main
contribution to $p_{0}$ and $P_{i03}$ gives the case when the true
message $\mbox{\boldmath $x$}_{1}'$ is selected, but
$\mbox{\boldmath $x$}_{i}'$ is not. Maximum of $p_{0}$
and $P_{i03}$ is attained when
${\mbox{\boldmath $x$}'}^{(1)t} = {\mbox{\boldmath $x$}_{1}'},
{\mbox{\boldmath $x$}'}^{(2)t} = {\mbox{\boldmath $x$}_{2}'}$,
${\mbox{\boldmath $x$}'}^{(3)t} = {\mbox{\boldmath $x$}_{3}'}$ and
$i \geq 4$. Moreover, we need
$d_{1}^{(t)} \leq d_{2}^{(t)} \leq d_{3}^{(t)} \leq d_{i}^{(t)}$ and
$d_{i}^{(t)} \geq d_{3}^{(t)} + 2A_{1}\tau_{3}$. Then neglecting
$\tau_{2}$, similarly to \eqref{S2a}--\eqref{S2b} we have
$$
\begin{gathered}
p_{0}(\mbox{\boldmath $\xi$}') \leq M^{4}{\mathbf P}\left(
(\mbox{\boldmath$x$}_{2}' - \mbox{\boldmath$x$}_{i}',
\mbox{\boldmath $\xi$}' + \sigma \mbox{\boldmath $\eta$}')
\geq A_{1}\tau_{3},
(\mbox{\boldmath$x$}_{3}' - \mbox{\boldmath$x$}_{i}',
\mbox{\boldmath $\xi$}' + \sigma \mbox{\boldmath $\eta$}')
\geq A_{1}\tau_{3}\Big|\{\xi_{i}'\}\right) \leq \\
\leq M^{4}{\mathbf P}\left(
(\mbox{\boldmath$x$}_{2}' + \mbox{\boldmath$x$}_{3}' -
2\mbox{\boldmath$x$}_{i}',
\mbox{\boldmath $\xi$}' + \sigma \mbox{\boldmath $\eta$}')
\geq 2A_{1}\tau_{3}\Big|\{\xi_{i}'\}\right) = \\
= M^{4}{\mathbf P}\left(
\sigma(\mbox{\boldmath$x$}_{2}' + \mbox{\boldmath$x$}_{3}' -
2\mbox{\boldmath$x$}_{i}',\mbox{\boldmath $\eta$}') \geq
2A_{1}\tau_{3} + 2\sqrt{A_{1}}\xi_{i}' -
\sqrt{A_{1}}\xi_{2}' - \sqrt{A_{1}}\xi_{3}'
\Big|\{\xi_{i}'\}\right) \leq \\
\leq M^{4}\exp\left\{-\frac{(2\sqrt{A_{1}}\tau_{3}-\xi_{2}' -
\xi_{3}' + 2\xi_{i}')_{+}^{2}}{6\sigma^{2}}\right\}
\end{gathered}
$$
and therefore (here $\xi, \zeta$ -- independent
${\cal N}(0,1)$-Gaussian random variables)
$$
\begin{gathered}
P_{i03} \leq e^{o(A_{1})}
{\mathbf P}\bigg\{\sqrt{A_{1}}(\xi_{i}' - \xi_{1}') +
\sqrt{2A_{2}}\xi'' -\frac{(2\sqrt{A_{1}}\tau_{3}-\xi_{2}' -\xi_{3}' +
2\xi_{i}')^{2}}{6\sigma^{2}}\geq A_{1}+ A_{2}\bigg\} \leq \\
\leq e^{o(A_{1})}{\mathbf P}\bigg\{\sqrt{2A_{1}/3}\zeta +
\sqrt{A_{1}/3}\zeta_{1}-\sqrt{A_{1}}\xi_{1}' + \sqrt{2A_{2}}\xi'' -
\frac{(2\sqrt{A_{1}}\tau_{3} + \sqrt{6}\zeta)^{2}}
{6\sigma^{2}}\geq A_{1}+ A_{2}\bigg\} = \\
= e^{o(A_{1})}{\mathbf P}\left\{\sqrt{2A_{1}}\zeta +
\sqrt{2(2A_{1}+3A_{2})}\xi
-\frac{\sqrt{3}(\sqrt{2A_{1}/3}\tau_{3} + \zeta)_{+}^{2}}
{\sigma^{2}} \geq \sqrt{3}(A_{1}+ A_{2})\right\},
\end{gathered}
$$
where we used the representations
$2\xi_{i}' -\xi_{2}' - \xi_{3}' = \sqrt{6}\zeta$,
$\xi_{i}' = \sqrt{2/3}\,\zeta + \zeta_{1}/\sqrt{3}$,
$\zeta \bot \zeta_{1}$ and similar ones.

Denoting $y\sqrt{A_{1}} = \zeta$, $x\sqrt{A_{1}} = \xi$ and
using the formula \eqref{Phi2}, similarly to \eqref{Pi01} we have
$$
\begin{gathered}
-2\ln P_{i03} \geq A_{1}\inf_{(x,y) \in {\cal A}}
\left(x^{2} + y^{2}\right) + o(A_{1}),  \\
{\cal A} = \left\{x,y: x - \varepsilon (y + a)^{2} \geq B \right\},
\end{gathered}
$$
where we omitted the sign of the positive part (similarly to
\eqref{Pi02}) and where
$$
\begin{gathered}
\varepsilon = \frac{1}{\sigma^{2}}\sqrt{\frac{3}{2(2+3\beta)}}, \qquad
a = \frac{2\tau_{3}-\sigma^{2}}{\sqrt{6}}, \qquad
B = \frac{6(1+\beta) + 4\tau_{3} - \sigma^{2}}
{2\sqrt{6(2+3\beta)}}.
\end{gathered}
$$
Now similarly to the case $P_{i02}$ we get as $A_{1} \to \infty$
\begin{equation}\label{Pi033}
\begin{gathered}
-\ln P_{i03} \geq \\
\geq \frac{A_{1}}{6}\left\{\frac{[6(1+\beta) +4\tau_{3} -
\sigma^{2}]^{2}}{8(2+3\beta)} +\frac{(2\tau_{3}-\sigma^{2})^{2}}{2} -
\frac{(2\tau_{3}-\sigma^{2})^{2}(2+3\beta)\sigma^{2}}
{6(1+\beta) + 4\tau_{3} -\sigma^{2}} + o(1)\right\}.
\end{gathered}
\end{equation}
For $\sigma^{2} \leq 1$ the formula \eqref{Pi033} can be simplified
as follows
\begin{equation}\label{Pi033sim}
\begin{gathered}
-\ln P_{i03} \geq \frac{A_{1}(1+\beta)(1 - \sigma^{2})}{4(2+3\beta)}
[2+3\beta + (1+2\tau_{3})^{2}].
\end{gathered}
\end{equation}

Consider $P_{i04}$. Maximum of $p_{0}$ and $P_{i04}$ is attained when
${\mbox{\boldmath $x$}'}^{(j)t} = {\mbox{\boldmath $x$}_{j}'}$,
$j = 1,\ldots,4$ and $i \geq 5$. Moreover, we need
$d_{1}^{(t)} \leq d_{2}^{(t)} \leq
d_{3}^{(t)} \leq d_{4}^{(t)} \leq d_{i}^{(t)}$.
Then for any $\sigma$ and $\beta \leq 1/2$
\begin{equation}\label{Pi04}
\begin{gathered}
P_{i04} \leq e^{o(A_{1})}
{\mathbf P}\left\{\sqrt{A_{1}}(\xi_{i}' - \xi_{1}') +
\sqrt{2A_{2}}\xi'' \geq A_{1}+ A_{2}, \xi_{2}' \geq \xi_{i}',
\xi_{3}' \geq \xi_{i}', \xi_{4}' \geq \xi_{i}' \right\} \leq \\
\leq e^{o(A_{1})}{\mathbf P}\left\{\sqrt{A_{1}}\xi_{i}' +
\sqrt{A_{1}+2A_{2}}\xi \geq A_{1}+ A_{2},
\xi_{2} \geq \sqrt{3}\xi_{i}'\right\} \leq \\
\leq e^{o(A_{1})} \min_{a \geq 0}{\mathbf P}\left\{
\sqrt{A_{1}}(1-a\sqrt{3})\xi_{i}' + a\sqrt{A_{1}}\xi_{2} +
\sqrt{A_{1}+2A_{2}}\xi \geq A_{1}+ A_{2}\right\} = \\
= e^{o(A_{1})}\min_{a \geq 0}{\mathbf P}\left\{
\sqrt{A_{1}[(1-a\sqrt{3})^{2} + a^{2} +1] + 2A_{2}} \xi \geq
A_{1}+ A_{2}\right\} \leq \\
\leq e^{o(A_{1})}{\mathbf P}\left\{\sqrt{5A_{1}/4+ 2A_{2}} \xi \geq
A_{1}+ A_{2}\right\} \leq \\
\leq \exp\left\{-\frac{2(1+\beta)^{2}A_{1}}
{5+8\beta} + o(A_{1})\right\} \leq e^{-(1+\beta)A_{1}/3 + o(A_{1})}.
\end{gathered}
\end{equation}
Therefore for $\sigma^{2} \leq 1$ and $\tau_{2} \leq 4/9$ we get from
\eqref{Pi02sim}, \eqref{Pi033sim} and \eqref{Pi04}
\begin{equation}\label{Pi03a}
\begin{gathered}
-\ln P_{i0} \geq \frac{(1+\beta)A_{1}}{4}\left[1 + \min\left\{
\frac{(1+ 2\tau_{2})^{2}}{3+4\beta},\frac{(1+2\tau_{3})^{2}}
{2+3\beta},1/3\right\}\right](1 - \sigma^{2}) + o(A_{1}).
\end{gathered}
\end{equation}
Note that if $\tau_{2},\tau_{3}$ satisfy conditions
\begin{equation}\label{Pi0cond}
\begin{gathered}
\tau_{2} \geq \frac{1}{\sqrt{15} +3} \approx 0.1455,
\qquad \tau_{3} \geq \frac{1}{2(\sqrt{42} + 6)}  \approx 0.04006,
\end{gathered}
\end{equation}
then for any $\beta \leq 1/2$ the formula \eqref{Pi03a} takes the
form
\begin{equation}\label{Pi0cond1}
\begin{gathered}
-\ln P_{i0} \geq \frac{A_{1}(1+\beta)(1 - \sigma^{2})}{3} + o(A_{1}).
\end{gathered}
\end{equation}

{\bf 3. Case $\mathbf{k = 3}$}. $\delta_{2} = 3A_{2}$.
This case takes place if the group
${\cal S}^{3}$  of three messages was selected and
$\mbox{\boldmath $x$}_{1}',\mbox{\boldmath $x$}_{i}' \in
{\cal S}^{3}$. Then
$\|\mbox{\boldmath $x$}_{i}'' -\mbox{\boldmath $x$}_{1}''\|= 3A_{2}$.
Main contribution to $p_{3}(\mbox{\boldmath $y$}')$ and $P_{i3}$
is given by case
$\left\{\mbox{\boldmath $x$}_{1}',\mbox{\boldmath $x$}_{i}'\right\} =
\left\{{\mbox{\boldmath $x$}'}^{(1)t},
{\mbox{\boldmath $x$}'}^{(2)t}\right\}$. Moreover, since we are
interested in the probability
${\mathbf P}\left\{X_{i} + Y_{i} \geq 0
\big|\mbox{\boldmath $y$}',\theta_{1}\right\}$ and
$\|\mbox{\boldmath $x$}_{i}'' -\mbox{\boldmath $x$}_{1}''\|/A_{2} = 3
> \|\mbox{\boldmath $x$}_{i}' -\mbox{\boldmath $x$}_{1}'\|/A_{1}= 2$,
then we may assume that $d_{1}^{(t)} \geq d_{2}^{(t)}$. More exactly,
without loss of generality, we may assume that $i = 2$ and
$\mbox{\boldmath $x$}_{i}' = \mbox{\boldmath $x$}_{2}' =
{\mbox{\boldmath $x$}'}^{(1)t}$,
$\mbox{\boldmath $x$}_{1}' = {\mbox{\boldmath $x$}'}^{(2)t}$,
$\mbox{\boldmath $x$}_{3}' = {\mbox{\boldmath $x$}'}^{(3)t}$. Then
first we have
$$
\begin{gathered}
P_{i3} \leq M^{3}{\mathbf P}\{d_{1} - d_{2} \geq 0,
d_{3}^{(t)} - d_{1}^{(t)} < 2A_{1}\tau_{2}|\theta_{1}\} \leq \\
\leq M^{3}{\mathbf P}\{
(\mbox{\boldmath $x$}_{2}' -\mbox{\boldmath $x$}_{1}',
\mbox{\boldmath $\xi$}') + \sqrt{3A_{2}}\xi''\geq A_{1}+ 3A_{2}/2,
\left(\mbox{\boldmath $x$}_{3}' -\mbox{\boldmath$x$}_{1}',
\mbox{\boldmath $\xi$}' + \sigma \mbox{\boldmath $\eta$}'\right)
\geq A_{1}(1-\tau_{2})|\theta_{1}\} \leq \\
= M^{3}{\mathbf P}\{\xi_{2}' - \xi_{1}' + \sqrt{3\beta}\xi'' \geq
\sqrt{A_{1}}(1+ 3\beta/2),\xi_{3}' - \xi_{1}' - \sigma \sqrt{2}\eta
\geq \sqrt{A_{1}}(1-\tau_{2})\}.
\end{gathered}
$$
Since
$\xi_{2}' - \xi_{1}' + \sqrt{3\beta}\xi'' \sim \sqrt{2+3\beta}\,\xi$
and $\xi_{3}' - \xi_{1}' - \sigma \sqrt{2}\eta \sim
\sqrt{2(1+\sigma^{2})}\,\zeta$, where $\xi, \zeta$ --
${\cal N}(0,1)$-Gaussian random variables with
${\mathbf E}(\xi \zeta) = -1/\sqrt{2(2+3\beta)(1+\sigma^{2})}$, then
using the inequalities \eqref{lem1} and \eqref{Phi1}, we get
as $A_{1} \to \infty$
$$
\begin{gathered}
-\ln P_{i3} \geq -\ln {\mathbf P}\left\{\xi\geq \frac{1}{2}
\sqrt{(2+3\beta)A_{1}}, \eta \geq (1-\tau_{2})
\sqrt{A_{1}/[2(1+\sigma^{2})]}\right\} + o(A_{1}) \geq \\
\geq \frac{A_{1}}{4}\left[\frac{2+3\beta}{2} + \frac{(1-\tau_{2})
+ (1-\tau_{2})^{2}}{1+\sigma^{2}}\right] + o(A_{1}) \geq \\
\geq \frac{A_{1}(1+\beta)}{4}\left[1 + \frac{\beta}{2(1+\beta)} +
\frac{2-3\tau_{2}}{(1+\beta)(1+\sigma^{2})}\right] + o(A_{1}).
\end{gathered}
$$
We limits ourselves only to values $\beta \leq 1/2$,
$\tau_{2} \leq 1/3$, $\sigma^{2} \leq 1$. Then
\begin{equation}\label{Pi3a}
\begin{gathered}
-\ln P_{i3} \geq \frac{A_{1}(1+\beta)}{3} + o(A_{1}).
\end{gathered}
\end{equation}

{\bf 4. Case $\mathbf{k = 4}$}. $\delta_{4} = 8A_{2}/3$. Similarly
to ${\cal S}_{3}$ maximum of $p_{0}$ and $P_{i4}$ is attained when
${\mbox{\boldmath $x$}'}^{(1)t} = {\mbox{\boldmath $x$}_{i}'},
{\mbox{\boldmath $x$}'}^{(2)t} = {\mbox{\boldmath $x$}_{1}'}$,
${\mbox{\boldmath $x$}'}^{(3)t} = {\mbox{\boldmath $x$}_{3}'}$,
${\mbox{\boldmath $x$}'}^{(4)t} = {\mbox{\boldmath $x$}_{4}'}$.
Moreover, we need $d_{i}^{(t)} \leq d_{1}^{(t)} \leq
d_{3}^{(t)} \leq d_{4}^{(t)}$ and
$d_{3}^{(t)} - d_{1}^{(t)} \leq 2A_{1}\tau_{2}$,
$d_{4}^{(t)} - d_{3}^{(t)} \leq 2A_{1}\tau_{3}$. Then we have
$$
\begin{gathered}
P_{i4} \leq M^{4}{\mathbf P}\{
(\mbox{\boldmath $x$}_{2}' -\mbox{\boldmath $x$}_{1}',
\mbox{\boldmath $\xi$}') +\sqrt{8A_{2}/3}\xi''\geq A_{1}+ 4A_{2}/3, \\
\left(\mbox{\boldmath $x$}_{3}' -\mbox{\boldmath$x$}_{1}',
\mbox{\boldmath $\xi$}' + \sigma \mbox{\boldmath $\eta$}'\right)
\geq A_{1}(1-\tau_{2}),
\left(\mbox{\boldmath $x$}_{4}' -\mbox{\boldmath$x$}_{3}',
\mbox{\boldmath $\xi$}' + \sigma \mbox{\boldmath $\eta$}'\right)
\geq -A_{1}\tau_{3}\} = \\
= M^{4}{\mathbf P}\Big\{\xi_{2}' - \xi_{1}' +
\sqrt{8\beta/3}\xi'' \geq \sqrt{A_{1}}(1+ 4\beta/3), \\
\xi_{3}' - \xi_{1}' + \sigma(\eta_{3}' - \eta_{1}') \geq
\sqrt{A_{1}}(1-\tau_{2}),
\xi_{4}' - \xi_{3}' + \sigma(\eta_{4}' - \eta_{3}')
\geq -\sqrt{A_{1}}\tau_{3}\Big\} = \\
= M^{4}{\mathbf P}\Big\{\sqrt{1 + 8\beta/3}\xi_{2} - \xi_{1}' \geq
\sqrt{A_{1}}(1+ 4\beta/3), \\
\sqrt{1+\sigma^{2}}\xi_{3} -\xi_{1}' - \sigma \eta_{1}' \geq
\sqrt{A_{1}}(1-\tau_{2}), \sqrt{1+\sigma^{2}}(\xi_{4} - \xi_{3})
\geq -\sqrt{A_{1}}\tau_{3}\Big\}
\end{gathered}
$$
and therefore
$$
\begin{gathered}
-2\ln P_{i4} \geq A_{1}\min_{\mbox{\boldmath $z$} \in {\cal A}}
\|\mbox{\boldmath $z$}\|^{2} + o(A_{1}), \qquad
\mbox{\boldmath $z$} = (z_{1},\ldots,z_{5}),  \\
{\cal A} = \Big\{\mbox{\boldmath $z$}:
\sqrt{1 + 8\beta/3}z_{2} - z_{1} \geq 1+ 4\beta/3,
\sqrt{1+\sigma^{2}}z_{3} -z_{1} - \sigma z_{5} \geq 1-\tau_{2}, \\
\sqrt{1+\sigma^{2}}(z_{4} - z_{3})\geq -\tau_{3}\Big\}.
\end{gathered}
$$
Minimum is attained when there are equalities in all three
inequalities. Then
$$
\begin{gathered}
z_{4} = z_{3} - \frac{\tau_{3}}{\sqrt{1+\sigma^{2}}}, \qquad
\sigma z_{5} = \sqrt{1+\sigma^{2}}z_{3} -z_{1} -1+\tau_{2},
\end{gathered}
$$
and after standard algebra we get
$$
\begin{gathered}
\min_{\mbox{\boldmath $z$} \in {\cal A}}
\|\mbox{\boldmath $z$}\|^{2} = \min_{z_{1},y_{3}}\Bigg\{z_{1}^{2} +
\frac{(3z_{1} + 3+4\beta)^{2}}{3(3+ 8\beta)} +
\frac{y_{3}^{2} + (y_{3} - \tau_{3})^{2}}{1+\sigma^{2}} +
\frac{(y_{3} -z_{1} -1+\tau_{2})^{2}}{\sigma^{2}}\Bigg\} = \\
= (1-\tau_{2})^{2} +\frac{(3\tau_{2} +4\beta)^{2}}{3(3+ 8\beta)}
+ \frac{\sigma^{2}\tau_{3}^{2}}{(1+\sigma^{2})(1+3\sigma^{2})}
- \dfrac{\left[1-\tau_{2}- \dfrac{3\tau_{2} +4\beta}{3+ 8\beta} -
\dfrac{\tau_{3}}{1+3\sigma^{2}}\right]^{2}}
{1 + \dfrac{3}{3+ 8\beta} + \dfrac{1}{\sigma^{2}} +
\dfrac{1+\sigma^{2}}{\sigma^{2}(1+3\sigma^{2})}} \geq \\
\geq (1-\tau_{2})^{2} +
\frac{(3\tau_{2} +4\beta)^{2}}{3(3+ 8\beta)} - \sigma^{2}
\left[1-\tau_{2}- \frac{3\tau_{2} +4\beta}{3+ 8\beta} -
\frac{\tau_{3}}{1+3\sigma^{2}}\right]^{2} \geq \\
\geq \left[(1-\tau_{2})^{2} +\frac{(3\tau_{2} +4\beta)^{2}}
{3(3+ 8\beta)}\right](1-\sigma^{2}) =
\frac{(3+4\beta)(3+4\beta - 6\tau_{2} + 6\tau_{2}^{2})(1-\sigma^{2})}
{3(3+8\beta)},
\end{gathered}
$$
since for $\tau_{2} + \tau_{3} \leq 1$ we have
$$
(1-\tau_{2})^{2} \geq \left[1-\tau_{2}- \frac{3\tau_{2} +4\beta}
{3+ 8\beta} - \frac{\tau_{3}}{1+3\sigma^{2}}\right]^{2}.
$$
Therefore if $\tau_{2} + \tau_{3} \leq 1$, then
\begin{equation}\label{Pi4}
\begin{gathered}
-\ln P_{i4} \geq
A_{1}f_{4}(\beta,\tau_{2})(1-\sigma^{2}) + o(A_{1}), \\
f_{4}(\beta,\tau_{2}) = \frac{(3+4\beta)(3+4\beta - 6\tau_{2} +
6\tau_{2}^{2})}{6(3+8\beta)}.
\end{gathered}
\end{equation}
Note that $f_{4}(1/2,\tau_{2}) \geq 1/2$, if
$\tau_{2} \leq (15 - \sqrt{105})/30 \approx 0.1584$.

Consider now the overall error probability $P_{i}$ from
\eqref{defpk1a}. Assuming
$\sigma^{2} \leq 1$, we set $\beta \leq 1/2$. Then for
$\tau_{2},\tau_{3}$ satisfying conditions \eqref{Pi0cond} and
$\tau_{2} \leq 1/3$ we get from \eqref{Pi0cond1}, \eqref{EPi2},
\eqref{Pi3a} and \eqref{Pi4} as $A_{1} \to \infty$
$$
\begin{gathered}
-\ln P_{i} \geq \min\{-\ln P_{ik}, k=0,2,3,4\} + o(A_{1}) \geq \\
\geq A_{1}(1 - \sigma^{2})\min\bigg\{\frac{(1+\beta)}{3},
\frac{1+2\beta}{4},f_{4}(\beta,\tau_{2})\bigg\} + o(A_{1}).
\end{gathered}
$$
We set $\beta = 1/2$. Then for any
$\left(\sqrt{5/3} - 1\right)/2 \approx 0.1455 \leq \tau_{2} \leq
(15 - \sqrt{105})/30 \approx 0.1584$ and
$\left(\sqrt{7/6} - 1\right)/2 \approx 0.04006 \leq \tau_{3} \leq
1-\tau_{2}$ we get as $A_{1} \to \infty$ (i.e. as $n \to \infty$)
\begin{equation}\label{Pi03a1}
\begin{gathered}
-\ln P_{i} \geq \frac{A_{1}(1 - \sigma^{2})}{2} + o(n) =
\frac{An(1 - \sigma^{2})}{3} + o(n).
\end{gathered}
\end{equation}
Since $P_{{\rm e}} \leq P_{{\rm e}1}$ (see \eqref{genPe}) and
$M = e^{o(n)}$, $n \to \infty$, then from \eqref{stransPe1},
\eqref{defpk1a} and \eqref{Pi03a1} we get
$$
\begin{gathered}
-\ln P_{{\rm e}} \geq \frac{An(1 - \sigma^{2})}{3} + o(n),
\end{gathered}
$$
which completes the proof of Theorem. \qquad $\Box$


\vspace{5mm}

\begin{flushleft}
{\small {\it Burnashev Marat Valievich} \\
Kharkevich Institute for Information Transmission Problems, \\
Russian Academy of Sciences, Moscow\\
 {\tt burn@iitp.ru}} \\
{\small {\it Yamamoto Hirosuke} \\
School of Frontier Sciences \\
The University of Tokyo, Japan \\
 {\tt hirosuke@ieee.org}}
\end{flushleft}%

\end{document}